\documentclass[aps,twocolumn,showpacs]{revtex4}
\usepackage{graphics}

\newcommand{\stac}[2]{\stackrel{\scriptscriptstyle {#1}}{#2}}

\begin{document}

\title{D-braneworld cosmology}

\author{Tetsuya Shiromizu\dag\ddag,
Takashi Torii\ddag \ and
Tomoko Uesugi\#}

\address{\dag\ Department of Physics, Tokyo Institute of Technology,
Tokyo 152-8551, Japan}

\address{\ddag\ Advanced Research Institute for Science and Engineering,
Waseda University, Tokyo 169-8555, Japan}

\address{\#\ Institute of Humanities and Sciences and Department of Physics, 
Ochanomizu University, Tokyo
112-8610, Japan}

\begin{abstract}
We discuss D-braneworld cosmology, that is, the brane is described by the
Born-Infeld action. Compared with the usual Randall-Sundrum braneworld 
cosmology
where the brane action is the Nambu-Goto one, we can see some
drastic changes at the very early universe: (i)universe may experience the
rapid accelerating phase
(ii)the closed universe may avoid the initial
singularity. We also briefly address the dynamics of the cosmology in the open
string metric, which might be  favorer than the induced metric from the
view point of the D-brane.

\end{abstract}



\maketitle

\section{Introduction}

One of the motivation for the braneworld is originated by the genius of a 
D-brane, that is,
open strings describing the Standard Model particles stick to the brane. So 
if one is serious
about this, we must employ the effective action for the D-brane, the 
Born-Infeld action, not
the Nambu-Goto membrane action\cite{BI}. In the presence of the matters on 
the brane,
the difference between the above two actions will appear.

The simplest braneworld model was proposed by Randall and 
Sundrum\cite{RSI,RSII} and
subsequently extended
to the cosmology by many peoples\cite{Tess,Roy,cosmos}.
However, the action for the brane is often assumed to be
the Nambu-Goto action and the action for the matters on the brane is simply 
added by
direct sum. In this paper we explore the cosmology on the D-brane (Related 
to the
tachyon condensation, the tachyon matter on the brane has been
also considered in the braneworld\cite{Mukohyama}).
As seen soon, our starting point is the 5-dimensional Einstein-Hilbert 
action plus the
Born-Infeld action (we call this D-braneworld). Then we consider the radiation
dominated universe on the D-brane. In this situation the brane action is 
described by Born-Infeld
one where the matter term is automatically included.

Here reminded that
there is a non-trivial aspect for the interpretation on the D-brane. 
According to Seiberg and
Witten\cite{SW}, the metric for the gauge field on the brane is given by 
$\stac{s}{g}_{\mu\nu}
=g_{\mu\nu}-(2\pi \alpha')^2(F^2)_{\mu\nu}$, not just the induced metric 
$g_{\mu\nu}$. $F$
represents the field strength of the gauge fields on the brane.
For this we will discuss the cosmology in the both
metric. See Ref.~\cite{Causal} for the causality issue.

Another motivation to think of the D-braneworld is the magnetogenesis in 
the very early universe
(See Ref.~\cite{PMF} for the comprehensive review and references).
The coherent magnetic field over the horizon scale currently has the limit
$B < 10^{-9}{\rm Gauss}$ from the cosmic microwave background 
anisotropy\cite{Silk}.
If such magnetic field with long coherent length actually exists, we must 
seek for
its primordial origin, especially in the inflating stage of the universe. 
It is well-known
that the magnetic field cannot be generated due to the conformal invariance of
the Maxwell theory in four dimension\cite{Parker}. But, the conformal 
invariance is broken due to
the non-linearity in the Born-Infeld
theory. Thus we might be able to have a magnetogenesis scenario during the 
inflation on the
D-braneworld.

The rest of this paper is organized as follows. In Sec.~\ref{Setup}, we 
describe the setup for
D-braneworld. We also comment on a holographic aspects in the D-braneworld.
In Sec.~\ref{Cosmological}, we will focus on the radiation dominated 
cosmology in the
induced metric.  First, we will average the energy-momentum tensor over the 
volume to
obtain the equation of state. And then we will see that the universe may be 
accelerating at the very
early stage and the closed universe could avoid the initial singularities.
In Sec.~\ref{Cosmology}, we briefly
reconsider the cosmology in the open string metric. In Sec.~\ref{Summary}, 
we give the summary and
discussion.  See Refs.~\cite{BIcosmos,BIcosmos2,BIBH}
for other issues on the Born-Infeld cosmology/black holes and 
Ref.~\cite{Gas} for
the D-brane effect in the different context.

\section{Setup}
\label{Setup}

The total action is composed of the bulk and the D-brane action:
%
\begin{eqnarray}
S=S_{\rm bulk} +S_{\rm BI},
\end{eqnarray}
%
where $S_{\rm bulk}$ is the 5-dimensional Einstein-Hilbert action with the 
negative
cosmological constant and $S_{\rm BI}$ is the Born-Infeld action for a D-brane:
%
\begin{eqnarray}
S_{\rm BI}=-\sigma \int d^4 x {\sqrt {-{\rm det}(g_{\mu\nu}+ (2\pi 
\alpha')F_{\mu\nu}) }},
\end{eqnarray}
%
where $F_{\mu\nu}$ is the field strength of the Maxwell theory on the 
brane. So $F_{\mu\nu}$ will
correspond to
the cosmic microwave background radiation if one thinks of the homogeneous 
and isotropic
universe as discussed later. In the Born-Infeld action the matter part is 
automatically
included. In the usual braneworld, on the other hand,
we assume that the action on the brane is given by the
Nambu-Goto action, $S_{\rm NG} \sim -\sigma \int d^4 {\sqrt{-g}}$, plus the 
matter
action $S_{\rm matter}$, in particular, $S_{\rm matter} \propto \int d^4 x 
{\sqrt {-g}}F^2$
for the Maxwell field on the brane.

In four dimensions $S_{\rm BI}$ becomes
%
\begin{eqnarray}
S_{\rm BI} & = & -\sigma \int d^4 x {\sqrt {-g}} \biggl[
1-\frac{1}{2}(2\pi\alpha')^2{\rm Tr}(F^2) \nonumber \\
  & & ~ +\frac{1}{8}(2\pi\alpha')^4 \bigl({\rm 
Tr}(F^2)\bigr)^2-\frac{1}{4}(2\pi\alpha')^4 {\rm
Tr}(F^4)\biggr]^{1/2}
\nonumber \\
& = & -\sigma \int d^4x {\sqrt {-g}}\biggl[1-\frac{1}{4}(2\pi\alpha')^2{\rm 
Tr}(F^2) \nonumber \\
& & ~+\frac{1}{32}(2\pi\alpha')^4\bigl({\rm 
Tr}(F^2)\bigr)^2-\frac{1}{8}(2\pi\alpha')^4{\rm
Tr}(F^4) \nonumber \\
& & +O(\alpha'^6) \biggr]. \label{BIaction}
\end{eqnarray}
%
{}From the first to the second line we expanded the square root and wrote 
down the
expression up to the order of $O(\alpha'^4)$. For the practice and 
simplicity, hereafter,
we employ this approximated action to discuss the {\it whole} history of 
cosmology.
Since the action for self-gravitationg D-brane is not still known, the present
treatment is conservative. The energy-momentum tensor on the brane is given by
%
\begin{eqnarray}
\stac{({\rm BI})}{T}_{\mu\nu}& = & -\sigma g_{\mu\nu}+4\pi \sigma
(2\pi \alpha')^2 \stac{({\rm em})}{T}_{\mu\nu} \nonumber \\
& & ~+\frac{1}{4}\sigma(2\pi \alpha')^4 {\rm Tr}(F^2) \biggl[ (F^2)_{\mu\nu}
-\frac{1}{8}g_{\mu\nu}{\rm Tr}(F^2) \biggr] \nonumber \\
& & ~-\sigma (2\pi\alpha')^4 \biggl[ 
(F^4)_{\mu\nu}-\frac{1}{8}g_{\mu\nu}{\rm Tr}(F^4)\biggr]
\nonumber \\
& =: & -\sigma g_{\mu\nu}+T_{\mu\nu},
\end{eqnarray}
%
where
%
\begin{eqnarray}
\stac{({\rm em})}{T}_{\mu\nu}=\frac{1}{4\pi} \biggl( F_\mu^{~\alpha} 
F_{\nu\alpha}
-\frac{1}{4}g_{\mu\nu}F_{\alpha\beta}F^{\alpha\beta}\biggr).
\end{eqnarray}
%
Hereafter we call $T_{\mu\nu}$ the energy-momentum tensor of the 
Born-Infeld matter.
To regard the above $ \stac{({\rm em})}{T}_{\mu\nu}$ as the energy-momentum 
tensor of
the usual Maxwell field on the brane, we set
%
\begin{eqnarray}
\sigma (2\pi\alpha')^2=1.
\end{eqnarray}
%
At the low energy limit $T_{\mu\nu}$ becomes
$T_{\mu\nu} \simeq -\sigma g_{\mu\nu}+\stac{({\rm em})}{T}_{\mu\nu}$ which 
is often
used in the usual braneworld.

Since we are interested in what happens on the brane, it is useful to 
consult with
the gravitational equation on the brane\cite{Tess} (See also \cite{Roy}):
%
\begin{eqnarray}
{}^{(4)}G_{\mu\nu}=8\pi G T_{\mu\nu}+\kappa^4 \pi_{\mu\nu}-E_{\mu\nu},
\end{eqnarray}
%
where
%
\begin{eqnarray}
8\pi G = \frac{\kappa^2}{\ell},\label{scale}
\end{eqnarray}
%
%
\begin{equation}
\pi_{\mu\nu} = -\frac{1}{4}T_{\mu\alpha}T^{\;\alpha}_{\nu} 
+\frac{1}{12}TT_{\mu\nu}
+\frac{1}{8}g_{\mu\nu}T^{\alpha}_{\;\beta} 
T^{\;\beta}_{\alpha}-\frac{1}{24}g_{\mu\nu}T^2
\end{equation}
%
and
%
\begin{eqnarray}
E_{\mu\nu}=C_{\mu\alpha\nu\beta}n^\alpha n^\beta.
\end{eqnarray}
%
In the above we supposed the Randall-Sundrum fine-tuning, that is,
%
\begin{eqnarray}
\frac{1}{\ell}=\frac{\kappa^2}{6}\sigma,
\end{eqnarray}
%
where $\ell$ is the curvature length of the five dimensional anti-deSitter 
spacetime.
Under this tuning, the net-cosmological constant on the brane vanishes.
We stress, however, that the tuning is not necessary for the discussion in 
this paper.

In general the above system is not closed on the brane except for the 
homogeneous-isotropic
universe due to the presence of
a part of the five dimensional Weyl tensor $E_{\mu\nu}$.
In the weak field limit, we can check
that the four dimensional Einstein gravity can be recovered\cite{Tama}.

Finally, it is worth noting that the trace-part of $\pi_{\mu\nu}$ is 
related to the trace part
of $T_{\mu\nu}$ as
%
\begin{eqnarray}
-\frac{\kappa^4}{3}\pi^\mu_{\;\mu} = \frac{\kappa^2}{\ell}T^\mu_{\;\mu} =
-\frac{\kappa^4}{12}\biggl[{\rm Tr}(F^4)-\frac{1}{4}\Bigl({\rm 
Tr}(F^2)\Bigr)^2 \biggr].
\end{eqnarray}
%
This is a realisation of the holography in the braneworld, that is,
$\pi^\mu_{\;\mu}$ represents a part of the quantum correction to the 
electromagnetic field
theory.  This is  because the Born-Infeld theory is a sort of 
phenomenological one for the
quantum  electrodynamics\cite{Hei}. In some cases we can show that 
$\pi^\mu_{\;\mu}$ is
identical  to the trace-anomaly of the quantum field theory on the 
brane\cite{holo1,holo2}.


\section{Cosmological models}
\label{Cosmological}

Let us focus on the homogeneous and isotropic universe. For simplicity, we 
consider the
single brane model. Then the metric on the brane is
%
\begin{eqnarray}
ds^2=-dt^2+a^2(t) \gamma_{ij}dx^i dx^j,
\end{eqnarray}
%
where $\gamma_{ij}$ is the metric of three dimensional unit sphere or unit 
hyperboloid
or flat space.
In this case we know $E^\mu_{\;\nu}=a^{-4}{\rm diag}(3\mu,-\mu,-\mu,-\mu)$ 
and $\mu$ is
proportional to the mass of the five-dimensional 
Schwarzschild-anti-deSitter spacetime
which is the bulk geometry. Moreover, the gravitational equation is closed 
on the brane,
that is, it is completely written in terms of the four dimensional 
quantities on the
brane. From now on we set $\mu=0$ which means that the bulk geometry
is exactly the anti-deSitter spacetime\footnote[1]{If the deviation
from the anti-deSitter or Schwarzschild-anti-deSitter spacetime
is, the gravitational equation is not
closed on the brane\cite{Tess}.}. Thus, the modified Friedmann equation becomes
%
\begin{eqnarray}
\label{friedman}
\biggl( \frac{\dot a}{a} \biggr)^2= \frac{\kappa^2}{3 \ell}\rho_{\rm BI}+
\frac{\kappa^4}{36}\rho_{\rm BI}^2-\frac{K}{a^2},
\end{eqnarray}
%
and
%
\begin{eqnarray}
\label{raychud}
\frac{\ddot a}{a}=-\frac{\kappa^2}{6\ell}(\rho_{\rm BI}+3P_{\rm 
BI})-\frac{\kappa^4}{36}
\rho_{\rm BI}(2\rho_{\rm BI}+3P_{\rm BI}).
\end{eqnarray}
%

By defining the electric and magnetic fields by
%
\begin{eqnarray}
E^i=F_0^{~i},~~~{\rm and}~~~B^i=\frac{1}{2}\epsilon^{ijk}F_{jk},
\end{eqnarray}
%
$T_{\mu\nu}$ can be rewritten as
%
\begin{eqnarray}
T_{00} & = & \frac{1}{2}(E^2+B^2)+\frac{\kappa^2\ell}{12}(E_i
B^i)^2 \nonumber \\
& & +\frac{\kappa^2\ell}{12}(E^2-B^2)\biggl[E^2-\frac{1}{4}(E^2-B^2 )\biggr],
\end{eqnarray}
%
%
\begin{eqnarray}
T_{0i}=\epsilon_{ijk}E^jB^k \biggl[1+\frac{\kappa^2\ell}{12}(E^2-B^2) \biggr],
\end{eqnarray}
%
and
%
\begin{eqnarray}
T_{ij} & = & -\biggl[ E_i E_j +B_i B_j -\frac{1}{2}g_{ij}(E^2+B^2) \biggr] 
\nonumber \\
& & ~~-\frac{\kappa^2\ell}{12}(E^2-B^2)\biggl[ E_i E_j +B_i B_j -B^2
g_{ij} \nonumber \\
& & -\frac{1}{4}g_{ij}(E^2-B^2)
\biggr]-\frac{\kappa^2\ell}{12}g_{ij}(E_k B^k )^2.
\end{eqnarray}
%

Since we identify the Maxwell field as the background radiation,
the energy density and the pressure of the Born-Infeld matter should be 
evaluated by averaging
over volume as\footnote[2]{If we consider full order terms of $\alpha'$,
the avegared energy-momentum tensor will be written by infinite number of terms
and cannot be represented by elementary functions. This is because
averaging is not commutative with the expansion in the operation.
As stressed before we will not consider the whole terms because we do not know
the action for the self-gravitationg D-brane. The action which we know is
only for probe one.}
%
\begin{eqnarray}
\rho_{\rm BI}& := & \langle T_{00} \rangle \nonumber \\
& = & \frac{1}{2}(E^2+B^2)+\frac{\kappa^2\ell}{36}B^2 E^2 \nonumber \\
& & +\frac{\kappa^2\ell}{12}(E^2-B^2)\biggl[ E^2-\frac{1}{4}(E^2-B^2)\biggr],
\end{eqnarray}
%
and
%
\begin{eqnarray}
P_{\rm BI}& := &\frac{1}{3}\langle T^i_i \rangle \nonumber \\
& = &
\frac{1}{6}(E^2+B^2)-\frac{\kappa^2\ell}{36}B^2 E^2 \nonumber \\
& & -\frac{\kappa^2\ell}{144}(E^2-B^2)(
E^2-5B^2 ).
\end{eqnarray}
%
We assumed that the $E_i$ and $B_i$ are random fields and the
coherent lengths are much shorter than the cosmological horizon
scales. In the above we assumed $\langle E_i E_j \rangle = (1/3)g_{ij}E^2$,
$\langle B_i B_j \rangle = (1/3)g_{ij}B^2$, $\langle E_i \rangle = \langle 
B_i \rangle =0$
and $\langle E_i B_j \rangle =0$ which are natural in the homogeneous and 
isotropic universe.
In addition, it is natural to assume 
``equipartition"\footnote[3]{Propery speaking, we must confirm this 
following the process to 
the equilibrium state based on the Boltzmann-like equation.}
\footnote[4]{If $F_{\mu\nu}$ corresponds to the primordial magnetic field,
it is natural to assume $E^2=0$ and $B^2 =2\epsilon \neq 0$. In this situation,
the energy density and pressure are given by
$\rho_{\rm BI}=\epsilon-\frac{\kappa^2\ell}{12}\epsilon^2$ and $P_{\rm BI}=
\frac{1}{3}\epsilon -\frac{5\kappa^2\ell}{36}\epsilon^2$. In 
Ref.~\cite{BIcosmos}
similar Born-Infeld fluid has been considered, but the authors
did not considered the D-braneworld. Just cosmology with the non-linear 
Maxwell field.}:
%
\begin{eqnarray}
E^2(t)=B^2(t)=:\epsilon.
\end{eqnarray}
%
Thus the energy density and the pressure are simply given by
%
\begin{eqnarray}
\label{eq-rho}
\rho_{\rm BI}=\epsilon + \frac{\kappa^2\ell}{36}\epsilon^2,
\end{eqnarray}
%
and
%
\begin{eqnarray}
\label{eq-p}
P_{\rm BI}=\frac{1}{3}\epsilon - \frac{\kappa^2\ell}{36}\epsilon^2.
\end{eqnarray}
%
By combining Eqs.~(\ref{eq-rho}) and (\ref{eq-p}), we obtain the equation 
of state
%
\begin{eqnarray}
\label{eos}
P_{\rm BI}=(\gamma_{BI}-1)\rho_{BI},
\end{eqnarray}
%
with an effective adiabatic index
%
\begin{eqnarray}
\gamma_{BI}=\frac{4}{3}\frac{1}{1+\frac{\kappa^2\ell}{36}\epsilon}.
\end{eqnarray}
%
At the low energy scale such $\epsilon \ll 36/\kappa^2\ell$,
the Born-Infeld matter behaves as just radiation fluid with 
$\gamma_{BI}\sim 4/3$ and
$\rho_{\rm BI} \sim (1/3)P_{\rm BI}$.
Note that the Born-Infeld matter looks like the
time-dependent cosmological constant if the second terms are dominated, 
i.e., $\gamma_{BI}\sim 0$.
 
\begin{figure}
\resizebox{3.0in}{!}{\includegraphics{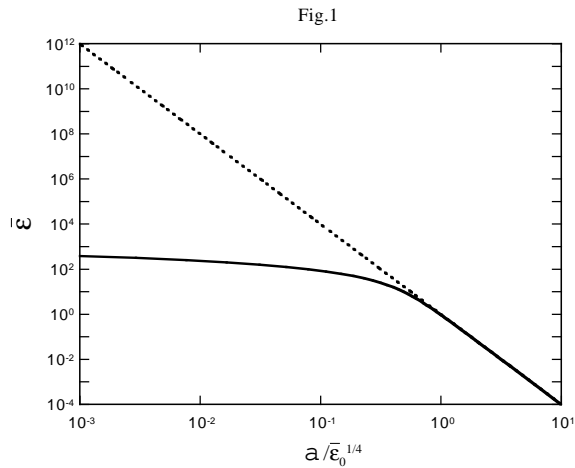}}
\caption{The log-plot of $\epsilon$. The horizontal axis is the scale 
factor. The dotted
line is the case of the ordinary radiation  fluid. The solid line is the 
case of the Born-Infeld
matter.}
\label{fig1}
\end{figure}

Now we have the equations of motion (\ref{friedman}) and (\ref{raychud}) 
and the equation of
state (\ref{eos}) for the Born-Infeld matter. It should be noted that these 
equations can be
scaled by defining new variables $\bar{t}:=t/\ell$,
$\bar{K}:=K\ell^2$, $\bar{\epsilon}:=\kappa^2\ell\epsilon$,
$\bar{\rho}_{BI}:=\kappa^2\ell\rho_{BI}$ and $\bar{P}_{BI}:=\kappa^2\ell 
P_{BI}$.
We can see the scale factor dependence of the energy density using the 
energy-conservation law,
$\dot \rho_{\rm BI}+3H(\rho_{\rm BI}+P_{\rm BI})=0$, on the  brane:
%
\begin{eqnarray}
\biggl( 1+\frac{\kappa^2\ell}{18} \epsilon \biggr) \dot \epsilon = -4H 
\epsilon.
\end{eqnarray}
%
It is easy to integrate the above equation and then
%
\begin{eqnarray}
\epsilon e^{\frac{\kappa^2\ell}{18}\epsilon}
=\frac{\epsilon_0}{a^4}.\label{state}
\end{eqnarray}
%
See Fig.~\ref{fig1} for $\epsilon$. Therein we also draw the ordinary 
radiation case
of $\epsilon \propto a^{-4}$. In the early universe, the Born-Infeld matter is
significantly suppressed  compared to the ordinary radiation fluid. We would
stress that the drastic changes from the ordinary braneworld come from
these features of the equation of state and the scale factor dependence of the
energy density $\epsilon$. For example, if one considers the
radiation dominated cosmology in the ordinary braneworld,
the correction terms to the Friedmann equation will be given
by $\pi^\mu_{\;\nu}=\frac{\rho^2}{12}(-1,1,1,1) $ and it will not play
role as a vaccum energy.

Here note that the matter does not always satisfy the ``strong energy
condition" $\rho+3P \geq 0$. For the D-brane matter,
%
\begin{eqnarray}
\rho_{\rm BI}+3P_{\rm BI}=2\epsilon-\frac{\kappa^2\ell}{18}\epsilon^2,
\end{eqnarray}
%
and
%
\begin{eqnarray}
2\rho_{\rm BI}+3P_{\rm BI}=3\epsilon-\frac{\kappa^2\ell}{36}\epsilon^2.
\end{eqnarray}
%
When $\gamma_{BI} < 2/3$, i.e., $\epsilon  > 36/\kappa^2\ell  \sim (10^3 
{\rm GeV})^4
(M_5/10^8{\rm GeV})^6$, the ``strong energy condition" is presumably
broken\footnote[5]{$M_5$ is the fundamental scale of five dimensional gravity 
and then
$M_5 \sim \kappa^{-1/3} \sim (M_{\rm pl}^2\ell^{-1})^{1/3}\sim 10^8(1{\rm 
mm}/\ell)^{1/3}{\rm GeV}$
in the single brane models. The string length becomes
$\ell_s = {\sqrt {\alpha'}} \sim \sigma^{-1/4} \sim
(\ell_{\rm pl} \ell)^{1/2} \sim 10^{-17} \times (\ell/1{\rm mm})^{1/2} {\rm 
cm}$}.
Thus the universe is accelerating
during such period, $\ddot a/a >0$. Furthermore, we have the opportunity that
the initial singularity can be avoided.

To see the quantitative feature of the dynamics of cosmology
it is useful to write down the generalized Friedmann equation
as usual:
%
\begin{eqnarray}
\dot a^2+V(a)=-K,
\end{eqnarray}
%
where
%
\begin{eqnarray}
V(a):=-\frac{\kappa^2}{3\ell}a^2 \rho_{\rm BI}
\biggl(1+\frac{\kappa^2\ell}{12}\rho_{\rm BI} \biggr).
\end{eqnarray}
%


\begin{figure}
\resizebox{3.0in}{!}{\includegraphics{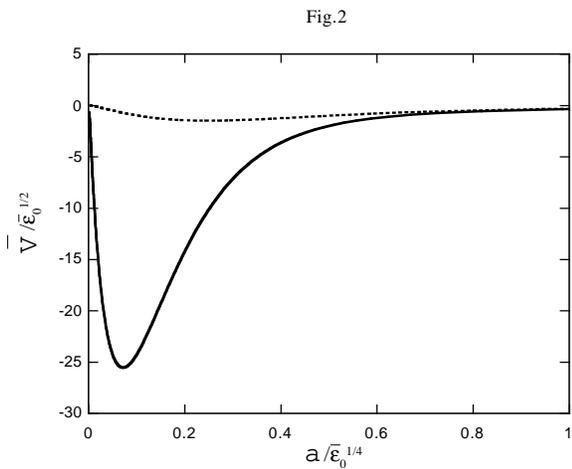}}
\caption{The potential profile of $\bar{V}(a):=V\ell^2$. The potential has 
the minimum and
vanishes at $a=0$.
We also plot the 4-dimensinal case with the BI matter (dashed line).}
\label{fig2}
\end{figure}

See Fig.~\ref{fig2} for the potential profile. Surprisingly, there is the 
minimum.
In addition, as shown analytically, the potential is zero at $a=0$.
Let us look at around $a=0$. $\epsilon$ can be approximately solved as
%
\begin{eqnarray}
\frac{\kappa^2 \ell}{18}\epsilon \sim -4 {\rm log}a.
\end{eqnarray}
%
We can see that the potential, indeed, is zero  at $a=0$ as seen in 
Fig.~\ref{fig2}. In the
current approximation, we see
%
\begin{eqnarray}
V(a) \sim -\frac{576}{\ell^2}a^2({\rm log}a)^4 \to  0, ~~~ (a\to 0).
\end{eqnarray}
%
As a result the closed universes is bounced around $a=0$! The flat or open 
universe
has the initial singularity. Near $a=0$, the behavior of the scale factor 
in the flat universe
becomes
%
\begin{eqnarray}
a(t) \sim e^{-\frac{\ell}{24t}},
\end{eqnarray}
%
and then $\ddot a /a \sim \frac{\ell^2}{576}\frac{1}{t^2} >0$, that is, the
universe is accelerating.
The bouncing behabior of the closed universe is owing to the existence of the
BI field as we can observe the similar form of the potential without the
quadratic terms $\pi^{\mu}_{\;\nu}$
in Fig.~\ref{fig2}. As we will see soon, however, the quantative behavior
of the scale factor is
quite different eqpecially around the bouncing, where the energy density
becomes large.

\begin{figure}
\resizebox{3.0in}{!}{\includegraphics{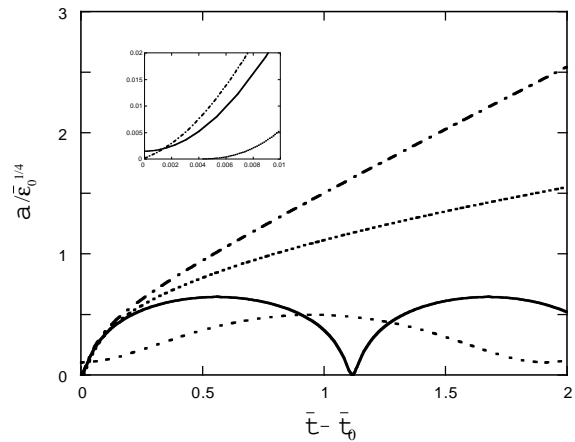}}
\caption{The behavior of the scale factor of the closed (solid line), flat 
(dashed line)
and open (dot-dashed line) cases in the D-braneworld.
We also plot the 4-dimensinal case with the BI matter for the closed
universe (dotted line).}
\label{fig3}
\end{figure}

For $\epsilon \ll 36/\kappa^2\ell$, as should be so, the universe is 
described by
the ordinary radiation dominated model. See Fig.~\ref{fig3} for the 
behavior of the scale factor.

\section{Cosmology in stringy view}
\label{Cosmology}

So far we investigated the D-braneworld in terms of the induced metric
$g_{\mu\nu}$. For the gauge field on the brane like photons, however,
the propagation of the field is described by the metric
%
\begin{eqnarray}
\stac{s}{g}_{\mu\nu}=g_{\mu\nu}-(2\pi \alpha')^2(F^2)_{\mu\nu}.
\end{eqnarray}
%
Hence it is fair to consider the gravitational
equation and cosmology in terms of $\stac{s}{g}_{\mu\nu}$. In the case of the
radiation dominated universe, the corresponding metric is given by
%
\begin{eqnarray}
d\tilde s^2 & = & \stac{s}{g}_{\mu\nu}dx^\mu dx^\nu \nonumber \\
& = & -d\tilde t^2+\tilde a(\tilde t)^2
\gamma_{ij}dx^i dx^j,
\nonumber \\
& = & -\biggl[1+(2\pi \alpha')^2 \langle (F^2)_{00} \rangle \biggr]dt^2
\nonumber \\
& & +\biggl[g_{ij}-(2\pi\alpha')^2 \langle (F^2)_{ij} \rangle \biggr]dx^i dx^j, 
\nonumber \\
& = & -\biggl( 1-\frac{\kappa^2\ell}{6}\epsilon \biggr)dt^2 \nonumber \\
& & +a^2
\biggl(1+\frac{\kappa^2\ell}{18}\epsilon \biggr) \gamma_{ij}dx^i dx^j,
\end{eqnarray}
%
where $\tilde t$ and $\tilde a (\tilde t)$ are the cosmic time and the 
scale factor
of the open string metric, respectively.
As discussed in Ref.~\cite{Causal}, the light-cone with respect to the open 
string
metric is smaller than that in the induced metric.
Let $n^\mu$ to be null vector for the induced metric. For the concreteness,
$n=\partial_t+(1/a)\partial_x$. Then
%
\begin{eqnarray}
\stac{s}{g}_{\mu\nu}n^\mu n^\nu & = & -(2\pi\alpha')^2(F^2)_{\mu\nu}n^\mu 
n^\nu,  \nonumber \\
& = & \frac{2\kappa^2\ell}{9}\epsilon >0,
\end{eqnarray}
%
and this means that $n$ is the spacelike in the open string metric.

We point out that the singularity appears at the {\it finite} value of
$\epsilon=\epsilon_c:= \frac{12}{\kappa^2\ell}$. The physical energy measured
also diverges because it is proportional to $1/{\sqrt {1-\kappa^2 \ell 
\epsilon /12}}$.
Anyway the universe evolves keeping
$\epsilon < \epsilon_c$. This means that there is no
drastic changes in the open string metric, but
slight modifications from the ordinary radiation dominated universe.

At first glance, we cannot examine the
interesting region where the feature of Born-Infeld becomes essential. However,
it might be
better to say that this is because of the limitation of the stringy metric.

\section{Summary and discussion}
\label{Summary}

  In this paper we have considered the Randall-Sundrum D-braneworld 
cosmology. Therein
the matter on the brane is described by the Born-Infeld action. As a first 
step, we
considered only the $U(1)$ gauge field and treated it as a sort of 
radiation fluids.
Then we examined the radiation dominated universe on the D-brane. We found that
the strong energy condition is broken in the very early stage and the 
universe is
more accelerated than the ordinary inflation. Furthermore the initial 
singularity is avoided
in closed universes. Thus we can conclude that we have the different 
history about the
early stage from the ordinary braneworld scenario if we are living on the 
D-brane, not on
the Nambu-Goto membrane.  In the acceleration phase, the ratio of the two 
scale factors at
different times is given by
$ a(t_f)/a(t_i) \sim e^{\frac{\kappa^2\ell}{72}(\epsilon_i -\epsilon_f)}$.
For the horizon problem $a(t_f)/a(t_i)> 10^{13} \times (100{\rm km}/s/{\rm 
Mpc}/H_0)
  ( T/10^3{\rm GeV})$ is required.
Then if $\frac{\kappa^2\ell}{72}\Delta \epsilon > 30 $ the horizon problem 
is solved.
However, the origin of the density fluctuation is not provided just in this 
scenario.

We should remark that we employed the approximated action in the second 
line of
Eq.~(\ref{BIaction}). We discussed the high energy regime where the
expansion is broken and the approximated
action may not be appropriate. Without
such approximation, however, we must treat the infinite series expansion 
due to the
volume averaging. Although our treatment contains this kind of problem, we 
could obtain
the important tendency. The future improvement for the treatment of the 
higher derivative
terms is desired.
 
We also addressed the D-braneworld in the open string metric, that is, {\it 
stringy view}.
As a result, the singularity appears just at the time when the non-trivial 
effects of the
D-brane are dominated. In this sense we might not be able to expect the 
significant
contribution from the D-braneworld specialties in the stringy view.

Since we obtained the nature that the Born-Infeld matter
contains vacuum/dark energy part, the current accelerating universe can be 
explained
in the D-braneworld without introducing the additional exotic fields like 
quitessence.

\section*{Acknowledgments}

TS would like to thank Kouji Hashimoto, Kei-ichi Maeda and Norisuke Sakai 
for fruitful
discussions.  To complete this work, the discussion during and after the 
YITP workshops
YITP-W-01-15 and  YITP-W-02-19 were useful. TS's work is supported by 
Grant-in-Aid for
Scientific Research from Ministry of Education, Science, Sports and Culture of
Japan (No. 13135208, No.14740155 and No.14102004).


\end{document}